\documentclass[aps,prl,twocolumn,superscriptaddress,floatfix,showpacs]{revtex4}

\usepackage{float}
\usepackage{placeins}
\usepackage{amssymb,amsmath}
\usepackage{graphicx}
\usepackage[colorlinks=true,linkcolor=blue,citecolor=blue,urlcolor=blue]{hyperref}

\begin{document}

\title{Application of a small oscillating magnetic field activates vortex motion deep inside the vortex-solid region of YBa$_2$Cu$_3$O$_7$}

\author{M.~Reibelt}
\email[]{reibelt@physik.uzh.ch}
%\author{S.~Weyeneth}
\affiliation{Physics Institute, University of Zurich, Winterthurerstrasse 190,
CH-8057 Zurich, Switzerland}
%\author{A.~Erb}
%\affiliation{Walther Meissner Institute, BAdW, D-85748 Garching, Germany}
%\author{A.~Schilling}
%\affiliation{Physics Institute, University of Zurich, Winterthurerstrasse 190,
%CH-8057 Zurich, Switzerland}
\date{\today}

\begin{abstract}
We have investigated the magnetic phase diagram of a fully oxygenated detwinned YBa$_2$Cu$_3$O$_7$ single crystal by means of a combination of a low-temperature thermal analysis (DTA) technique with a simultaneous application of a small oscillating magnetic field. We observed a dip-shaped feature in magneto-caloric data deep inside the vortex-solid region of YBa$_2$Cu$_3$O$_7$, which was activated by the oscillating magnetic field. The feature can be explained by a self-heating effect due to vortex motion. However, the origin of this new feature is unknown.
\end{abstract}


\maketitle

\section{Introduction}
The investigation of the magnetic phase diagram of high-$T_c$ superconductors has been of great importance since their discovery. The effects of oscillating magnetic fields on the vortex lattice of type-II superconductors have been investigated theoretically \cite{Brandt1986November,Brandt1994April,Brandt1996March,Brandt1996August,Mikitik2000September,Mikitik2001August,Brandt2002July,Brandt2004,Brandt2004a,Brandt2004b,Brandt2007}. Experimentally, the application of a small oscillating magnetic field (so-called shaking field) $h_{ac}$ has turned out to be a useful tool for the investigation of the properties of type-II superconductors \cite{Cape1968February,Risse1997June,Willemin1998September,Avraham2001May,Lortz2006September,Uksusman2009,Weyeneth2009,Reibelt2010March}. In this work we repeated the measurements of \cite{Reibelt2011September} but with an additional shaking field applied. We report on the, to the best of our knowledge, first observation of a dip-shaped feature in the magneto-caloric data $\Delta T(H)$ that appeared only in the field sweeps with simultaneously applied shaking field, while no such feature was present in the measurements without shaking field.

\section{Sample Characterization and experimental details}
The investigated sample (mass $\approx 6.5\, $mg) was an ultrapure YBa$_2$Cu$_3$O$_7$ single crystal \cite{Erb1996}. In order to reduce vortex pinning, the crystal was slightly overdoped to achieve full stoichiometry, and it was mechanically detwinned \cite{White2009March,Hinkov2004August}. The crystal has a critical temperature $T_c \approx 86\, $K and in a magnetic field of $7\, $T the melting of the vortex lattice was observed at $T_m(7T) \approx 80\, $K \cite{ForganPrivateCommunication}. The crystal was part of a mosaic of crystals which was studied in SANS experiments by White \emph{et al.}~\cite{White2009March,White2011}. We conducted measurements of the quasi-isothermal magneto-caloric effect on the crystal, which was aligned with its $c$-axis parallel to the external magnetic field ($H \parallel c$). The calorimetric measurements were done in a low-temperature home-built differential-thermal analysis (DTA) setup \cite{Schilling2007March}. Prior to a measurement, we cooled the sample to the target temperature in a constant magnetic field $\mu_0H = 7\, $T. The magneto-caloric effect was then monitored while decreasing $H$ to zero at a constant rate $\mu_0 dH/dt \approx -1\, $mTs$^{-1}$; the temperature varied less than $1\, $K for the whole field sweep. The setup and measurement procedure were the same as in our previous work on this crystal \cite{Reibelt2011September}, except that we simultaneously applied a shaking field $h_{ac}$ in addition to the main magnetic field $H$ during the measurement. The shaking field was applied perpendicular to the main magnetic field ($h_{ac} \perp H$).

\section{Results}
\FloatBarrier
\begin{figure}
\includegraphics[width=75mm,totalheight=200mm,keepaspectratio]{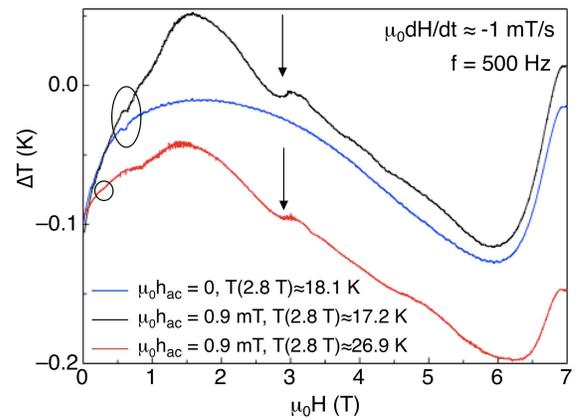}
\caption {Magneto-caloric data $\Delta T(H) = T_{ref}(H)-T_{sample}(H)$ of an YBa$_2$Cu$_3$O$_7$ single crystal for several field sweeps at different temperatures; with and without an applied shaking field. \label{fig.PhysicaC_2011_plot_1}}
\end{figure}
In Fig.~\ref{fig.PhysicaC_2011_plot_1} we plotted the magneto-caloric data $\Delta T(H) = T_{ref}(H)-T_{sample}(H)$ for various temperatures on ramping the magnetic field $H$ from $7\, $T down to zero. We labeled each curve with the sample temperature at $H=2.8\, T$, which is approximately the magnetic field where we observed the feature in the measurements with additionally applied shaking field. For the black ($T(2.8\, T)=17.2\, $K) and red ($T(2.8\, T)=26.9\, $K) curve we applied a shaking field during the measurement with amplitude $h_{ac} \approx 0.9\, $mT and frequency $f = 500\, $Hz. For comparison we also plotted a field sweep where no shaking field was applied (blue curve, $T(2.8\, T)=18.1\, $K). All plotted measurements ($h_{ac} = 0$ and $h_{ac} \neq 0$) exhibit the already in \cite{Reibelt2011September} reported first-order phase transition below $1\, $T, marked by ellipses. Strikingly, a second, broader dip-shaped feature sets in at around $2.8\, $T in the two measurements with applied shaking field (marked by arrows), while no such feature is present in the measurements without shaking field. The dip-shaped feature in the $\Delta T(H)$ curves results from a heating of the sample, which is most likely caused by a self-heating effect of the sample due to shaking field activated vortex motion. The vortex shaking in combination with the field sweep depins vortices which start to ``walk" through the sample \cite{Brandt2002July}, thereby dissipating energy which heats up the sample. We want to emphasize, that this feature occurs deep inside the vortex-solid region of YBa$_2$Cu$_3$O$_7$. However, although unlikely we can not rule out that the feature stems from the latent heat of a first-order phase transition, since we have no data with simultaneously applied shaking field on increasing the magnetic field or from a temperature sweep that should show a cooling effect in case of a first-order phase transition on crossing the supposed phase-transition line from the opposite side. Additional measurements are necessary to clarify this issue.\\
\FloatBarrier
\begin{figure}
\includegraphics[width=75mm,totalheight=200mm,keepaspectratio]{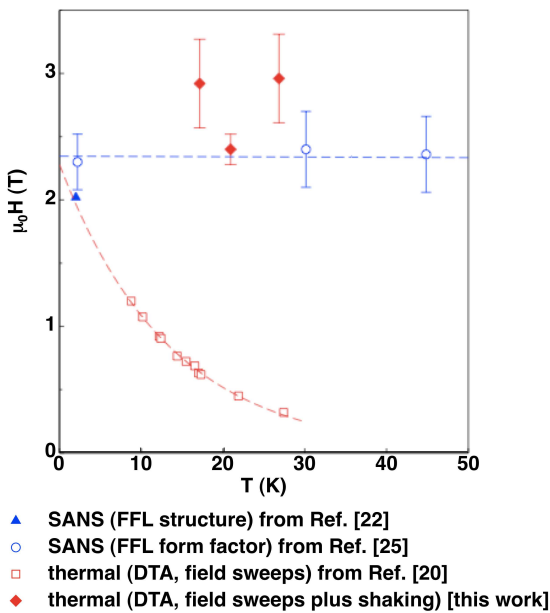}
\caption {Low-field part of the phase diagram of YBa$_2$Cu$_3$O$_7$ for $H \parallel c$. The position of the shaking field activated feature is drawn in as red diamonds. \label{fig.PhysicaC_2011_plot_2}}
\end{figure}
\FloatBarrier
\indent In Fig.~\ref{fig.PhysicaC_2011_plot_2} we replotted the low-field part of the phase diagram of YBa$_2$Cu$_3$O$_7$ from Fig.~2(a) of Ref.~\cite{Reibelt2011September}, which shows the phase-transition line for the first-order phase transition at low magnetic fields which we reported already in \cite{Reibelt2011September} (red dashed line in Fig.~\ref{fig.PhysicaC_2011_plot_2}). We added the data from the SANS measurements of White \emph{et al.}~\emph{et al.}~\cite{White2009March,White2011} (blue triangle and circles in Fig.~\ref{fig.PhysicaC_2011_plot_2}), who observed a vortex-lattice reorientation transition with a temperature independent phase-transition line (blue dashed line in Fig.~\ref{fig.PhysicaC_2011_plot_2}). Finally, we added to the magnetic phase diagram in Fig.~\ref{fig.PhysicaC_2011_plot_2} as red diamonds the data of the onset of the shaking field activated feature, which was marked with arrows in Fig.~\ref{fig.PhysicaC_2011_plot_1}. The data point at $\approx 20\, K$ stems from a field sweep which we did not show in Fig.~\ref{fig.PhysicaC_2011_plot_1}. For clarity reasons we omitted the error bars for the data from Refs.~\cite{White2009March,Reibelt2011September}.
The proximity of the
new feature to the vortex-lattice reorientation transition in the magnetic phase diagram may be coincidental.
White \emph{et al.}~\cite{White2009March} suggested that the vortex-lattice reorientation transition,
which they observed at about $2\, $T, is probably driven by Fermi surface effects. Decreasing the magnetic field
increases the distance between the vortices, which in turn lowers the influence of the non-local effects
which couple the vortex lattice to the Fermi surface. Vortex pinning should be affected by a symmetry
change of the vortex lattice \cite{Silhanek2001,Silhanek2002}. Approaching the vortex-lattice reorientation transition on
decreasing the magnetic field likely changes the vortex pinning potential already somewhat before the transition
is actually reached, thereby aiding the depinning process activated by the combined action of the shaking field
and the decreasing main magnetic field.

\section{Conclusions}
To conclude, we have investigated the magnetic phase diagram of a fully oxygenated
detwinned YBa$_2$Cu$_3$O$_7$ single crystal by means of a combination of a low-temperature
thermal analysis (DTA) technique with a simultaneous application of a shaking field.
We found a shaking field activated dip-shaped feature deep inside the vortex-solid region of YBa$_2$Cu$_3$O$_7$,
which can be explained by a self-heating effect due to activated vortex motion. Merely a few preliminary
measurements were performed with a shaking field simultaneously applied. The proximity of the
new feature to the vortex-lattice reorientation transition in the magnetic phase diagram may be coincidental, but
an influence of the reorientation transition on the vortex pinning potential is possible.
However, the origin of the in this work reported new feature is unknown.

\section{Acknowledgements}
We thank to the group of A.~Schilling, E.~M.~Forgan, J.~S.~White, and V.~Hinkov for their support and useful discussions, and we thank A.~Erb for providing the sample. This work was supported by the Schweizerische Nationalfonds zur F\"{o}rderung der Wissenschaftlichen Forschung, Grants No.~$20$-$111653$ and No.~$20$-$119793$.

\section{References}
%\bibliographystyle{jabrefformat}
%\bibliography{ReferenzenDatei}

\begin{thebibliography}{10}
\newcommand{\enquote}[1]{`#1'}

\bibitem{Brandt1986November}
E.~H. Brandt.
\newblock Phys. Rev. B \textbf{34}, 6514 (1986).

\bibitem{Brandt1994April}
E.~H. Brandt.
\newblock Phys. Rev. B \textbf{49}, 9024 (1994).

\bibitem{Brandt1996March}
E.~H. Brandt and A.~Gurevich.
\newblock Phys. Rev. Lett. \textbf{76}, 1723 (1996).

\bibitem{Brandt1996August}
E.~H. Brandt.
\newblock Phys. Rev. B \textbf{54}, 4246 (1996).

\bibitem{Mikitik2000September}
G.~P. Mikitik and E.~H. Brandt.
\newblock Phys. Rev. B \textbf{62}, 6800 (2000).

\bibitem{Mikitik2001August}
G.~P. Mikitik and E.~H. Brandt.
\newblock Phys. Rev. B \textbf{64}, 092502 (2001).

\bibitem{Brandt2002July}
E.~H. Brandt and G.~P. Mikitik.
\newblock Phys. Rev. Lett. \textbf{89}, 027002 (2002).

\bibitem{Brandt2004}
E.~H. Brandt and G.~P. Mikitik.
\newblock Supercond. Sci. Technol. \textbf{17}, S1 (2004).

\bibitem{Brandt2004a}
E.~H. Brandt and G.~P. Mikitik.
\newblock Physica C \textbf{404}, 69 (2004).

\bibitem{Brandt2004b}
E.~H. Brandt and G.~P. Mikitik.
\newblock Physica C \textbf{408-410}, 514 (2004).

\bibitem{Brandt2007}
E.~H. Brandt and G.~P. Mikitik.
\newblock Supercond. Sci. Technol. \textbf{20}, S111 (2007).

\bibitem{Cape1968February}
J.~A. Cape and I.~F. Silvera.
\newblock Phys. Rev. Lett. \textbf{20}, 326 (1968).

\bibitem{Risse1997June}
M.~P. Risse, \emph{et~al.}
\newblock Phys. Rev. B \textbf{55}, 15191 (1997).

\bibitem{Willemin1998September}
M.~Willemin, \emph{et~al.}
\newblock Phys. Rev. B \textbf{58}, R5940 (1998).

\bibitem{Avraham2001May}
N.~Avraham, \emph{et~al.}
\newblock Nature (London) \textbf{411}, 451 (2001).

\bibitem{Lortz2006September}
R.~Lortz, \emph{et~al.}
\newblock Phys. Rev. B \textbf{74}, 104502 (2006).

\bibitem{Uksusman2009}
A.~Uksusman, \emph{et~al.}
\newblock J. Appl. Phys. \textbf{105}, 093921 (2009).

\bibitem{Weyeneth2009}
S.~Weyeneth, \emph{et~al.}
\newblock J. Supercond. Nov. Magn. \textbf{22}, 325 (2009).

\bibitem{Reibelt2010March}
M.~Reibelt, \emph{et~al.}
\newblock Phys. Rev. B \textbf{81}, 094510 (2010).

\bibitem{Reibelt2011September}
M.~Reibelt, \emph{et~al.}
\newblock Supercond. Sci. Technol. \textbf{24}, 105019 (2011).

\bibitem{Erb1996}
A.~Erb, \emph{et~al.}
\newblock Physica C \textbf{258}, 9 (1996).

\bibitem{White2009March}
J.~S. White, \emph{et~al.}
\newblock Phys. Rev. Lett. \textbf{102}, 097001 (2009).

\bibitem{Hinkov2004August}
V.~Hinkov, \emph{et~al.}
\newblock Nature \textbf{430}, 650 (2004).

\bibitem{ForganPrivateCommunication}
Private communication with E.~M.~Forgan (2009).

\bibitem{White2011}
J.~S. White, \emph{et~al.}
\newblock Phys. Rev. B \textbf{84}, 104519 (2011).

\bibitem{Schilling2007March}
A.~Schilling and M.~Reibelt.
\newblock Rev. Sci. Instrum. \textbf{78}, 033904 (2007).

\bibitem{Silhanek2001}
A.~V. Silhanek, \emph{et~al.}
\newblock Phys. Rev. B \textbf{64}, 012512 (2001).

\bibitem{Silhanek2002}
A.~V. Silhanek, \emph{et~al.}
\newblock NATO Science Series (Kluwer Academic Publishers) \textbf{67}, 255 (2002).

\end{thebibliography}

\end{document}